# Observation of coherent perfect absorption at an exceptional point

Changqing Wang[1], William R. Sweeney[2,3], A. Douglas Stone[2,3], and Lan Yang[1]*

**Affiliations:**

[1]Department of Electrical and Systems Engineering, Washington University, St Louis, MO, 63130, USA.

[2]Departments of Applied Physics and Physics, Yale University, New Haven, CT, 06520, USA.

[3]Yale Quantum Institute, Yale University, New Haven, CT 06520, USA.

**\*** Correspondence to: yang@seas.wustl.edu

The past few years have witnessed growing interests in exceptional points (EPs) in various domains, including photonics, acoustics and electronics. However, EPs have mainly been realized based on the degeneracy of resonances of physical systems; distinct degeneracies occur relating to the absorption properties of waves, with distinct physical manifestations. Here we demonstrate this physically different kind of exceptional point, by engineering degeneracies in the absorption spectrum of optical microcavities with dissipation. We experimentally distinguish the conditions to realize a resonant EP and an absorbing EP. Furthermore, when the optical loss is optimized to achieve perfect absorption at such an EP, we observe an anomalously broadened lineshape in the absorption spectra, as predicted by theory. The distinct scattering properties enabled by this type of EP creates new opportunities for both the fundamental study and applications of non-Hermitian singularities.

The fast-growing field of non-Hermitian physics has created a fertile ground for studying unconventional physical phenomena in open systems (*1–3*). Among them, many of the most intriguing phenomena occur at exceptional points (EPs), singularities in the parameter space of a wave system at which two or more discrete eigenvalues and the associated eigenstates of a non-Hermitian system coalesce (*4*). While EPs can occur in any non-Hermitian linear system, the most familiar and well-studied experimental examples are EPs associated with the degeneracies of the resonance frequencies of wave systems in optics, quantum mechanics and acoustics, corresponding to the discrete, purely outgoing solutions of the relevant wave equations. Such a resonant EP, which generically occurs for complex frequencies, $\omega = \omega_r - i\gamma$, leads to interesting physical effects in its vicinity, as either parameters of the system or the probing frequency is varied. For example, as a parameter is swept through the EP, the eigenfrequency splitting of the two resonances passing through degeneracy has a parametrically enhanced sensitivity to small environmental perturbations, leading to EP enhanced sensing (*5–9*). In addition, the behavior of the eigenfrequencies in the vicinity of the EP exhibits non-analytic behavior and non-trivial topology, leading to phenomena such as asymmetric state transfer by encircling an EP (*10–12*), and the emergence of a hybrid topological invariant in the case of high order EPs (greater than double degeneracy) (*13*), as well as generation of bulk Fermi arcs connecting an EP pair (*14*). In addition, the collapse of the dimensionality of the eigenvector space at the EPs induces enhanced nonlinear optical/mechanical effects (*15–17*) and excess spontaneous emission noise (*18, 19*). Moreover, since the onset of laser emission corresponds to the non-generic case in which a resonance exists at a real frequency (*20*), an EP for two resonances engineered to have frequencies at or very near the real axis leads to unusual lasing behavior, including chiral laser emission (*21, 22*), laser linewidth broadening (*19, 23*), single-mode lasing (*24, 25*), loss induced laser revival (*26*), lasing/perfect absorption switching (*27*), etc. Except for the case of lasing, the resonant EPs involved are not at real frequency, and they are observed by their effects on scattering at real frequency, i.e., under quasi-steady-state excitation (*28–31*). Since the onset of lasing leads to a linear instability, which is stabilized by non-linear saturation, it is difficult experimentally to study the purely linear behavior near such a lasing EP. Here we study a different type of EP in an optical microcavity system, associated with perfect absorption, which has no instability at real frequency, and presents interesting EP-related behavior not previously observed.

As noted, resonances are associated with purely outgoing wave solutions (Sommerfeld or radiation boundary conditions); these are non-Hermitian boundary conditions which can only be solved at discrete, generically complex, frequencies, $\omega = \omega_r - i\gamma$. Similarly, purely incoming wave solutions are also non-Hermitian and will have solutions at such discrete complex frequencies. When the scattering structure is passive (no absorption or gain), these frequencies are just the complex conjugate of the resonance frequencies, $\omega = \omega_r + i\gamma$. In such passive structures neither the resonances nor the incoming solutions can exist at real frequency. Some time ago, it was shown that the analog of adding gain to achieve lasing was to add absorption to the same structure so as to make the incoming solution occur at real frequency (*32*). Such a structure is known as a Coherent Perfect Absorber (CPA); it is the time-reversed version of a laser at threshold and the input wavefront must be tuned to be the incoming version of the corresponding lasing mode to be fully absorbed (*33–36*). Other, non-optimized inputs will not be perfectly absorbed. However, unlike lasing/resonance, there is no intrinsic instability or divergence associated with a CPA. From the point of view of scattering, a resonance is a pole of the scattering matrix, S, where the response to an input diverges; whereas the incoming solutions correspond to zeros of S, where there is zero response to a finite input, specifically for the eigenvector of S with eigenvalue equal to zero.

Following the analogy to resonant EPs, it was recently proposed by two of the authors and collaborators (*37*) that with appropriate tuning, the eigenfrequencies of two incoming solutions of a wave equation could become degenerate, leading to what we term an EP Zero (when this happens at a complex frequency), and to CPA EP, when the system parameters are tuned so that this occurs at a real frequency. A number of interesting phenomena distinct from resonant EPs were predicted for such absorbing EPs, which we are able to observe and explore in the current work. A subtlety associated with EP zeros is that they are not observable in steady-state for lossless (non-absorbing) systems, as a distinct phenomenon from resonant EPs: since the zeros occur at the complex conjugate frequencies of the resonances, any steady-state scattering phenomenon associated with resonant EPs is also associated with EP zeros and there is no way to independently probe their properties in steady state. If absorption is present or added to the system, this symmetry is broken, and one can independently measure effects associated with zeros as opposed to poles, as we will show below. Specifically, we will demonstrate a distinct transition in the complex eigenfrequencies associated with the EP zeros from that associated with resonant EPs, as system parameters are varied. We will also probe and confirm the unique properties associated with a real EP zero, referred to as CPA EP, because in this case one observes EP behavior associated with perfect absorption of input signals. In this article it will be important to distinguish between absorption as a loss mechanism vs. scattering or coupling loss, which is non-dissipative. While increasing both types of loss causes resonances to move away from the real frequency axis, scattering loss does this for the zeros but absorption loss does the opposite.

A final general point of relevance is that resonant EPs have often been treated within temporal coupled mode theory (TCMT) (*38*), in terms of a non-Hermitian effective Hamiltonian, H$_{\text{eff}}$, the degeneracy of which corresponds to an EP. Typically, this Hamiltonian is used to construct the S-matrix of the system. In systems with parity, P, and time reversal symmetry, T, (no gain or absorption), or the frequently studied PT-symmetric system, with balanced gain or loss, an EP of H$_{\text{eff}}$ implies a corresponding EP of the S-matrix at the same system parameters (within the TCMT approximation). It does not seem to have been appreciated that in generic cases, without such symmetries, at EPs of the effective Hamiltonian (or more generally of the wave operator with appropriate boundary conditions), the S-matrix is not defective, i.e., is not at an EP (*37*). The scattering behavior at an absorbing EP is qualitatively different in the generic case (no symmetry) as compared to the case with symmetry, as we will demonstrate below. The generic case shows an anomalous quartic absorption lineshape, whereas the more symmetric case shows a quadratic absorption lineshape. Both cases are described within the TCMT model described in the next section.

**Theoretical model and experimental setup**

We study absorbing EPs experimentally in a coupled system of optical microcavities described below. The first theoretical treatment of absorbing EPs was given in (*37*). The properties we will explore in the current work were first derived there by considering a model of coupled standing-wave (SW) slab cavities with input and output channels on both sides (Fig. 1A) (*37*). The reflection and transmission amplitudes for such a two-channel scattering system, denoted by $r_1$ ($r_2$) and $t_1$ ($t_2$), respectively, lead to a 2 x 2 S matrix $S = \begin{pmatrix} r_1 & t_2 \\ t_1 & r_2 \end{pmatrix}$ (Fig. 1A). We realize an equivalent model with a different geometry via coupled WGM microcavities ($\mu R_1$ and $\mu R_2$) (Fig. 1B). They have resonant frequencies $\omega_{1,2}$ and intrinsic loss rates $\gamma_{1,2}$, and are coupled to each other with the coupling strength $\kappa$ and coupled to two single-mode taper fiber waveguides with the coupling strengths $\gamma_{c1}$ and $\gamma_{c2}$, respectively. The intrinsic losses are dominated by the absorption loss from the material and surface impurities, while the radiation loss is comparably small and negligible, consistent with the theoretical assumptions of (*37*), The incident laser light is injected from port 1 and 3, while the output signal spectra are monitored at port 2 and 4. While in general such a system has four ports and four scattering channels, due to the negligible backscattering of this structure, only the clockwise (CW) mode in $\mu R_1$ and the counterclockwise (CCW) mode in $\mu R_2$ are excited as long as only ports 1, 3 are used for input waves. The reflection, transmission amplitudes are defined as the amplitudes detected at ports 2, 4 when port 1 is excited and similarly for excitation of port 3 (Fig. 1B). Transmission is defined as passing through the coupled resonators and reflection as remaining in the same waveguide as the input wave. This setup can be described by TCMT and leads to an effective Hamiltonian (supplementary text S1):

$$H_{eff} = \begin{pmatrix} \omega_1 - i\frac{\gamma_1 + \gamma_{c1}}{2} & \kappa \\ \kappa & \omega_2 - i\frac{\gamma_2 + \gamma_{c2}}{2} \end{pmatrix}, \qquad (1)$$

and the corresponding 2-by-2 S matrix in the frequency domain is given by

$$S(\omega) = \begin{pmatrix} 1 - i\frac{\gamma_{c1}\Delta_2}{\Delta_1\Delta_2 - \kappa^2} & -i\frac{\sqrt{\gamma_{c1}\gamma_{c2}}\kappa}{\Delta_1\Delta_2 - \kappa^2} \\ -i\frac{\sqrt{\gamma_{c1}\gamma_{c2}}\kappa}{\Delta_1\Delta_2 - \kappa^2} & 1 - i\frac{\gamma_{c2}\Delta_1}{\Delta_1\Delta_2 - \kappa^2} \end{pmatrix}, \qquad (2)$$

where $\Delta_{1,2} = \omega - \omega_{1,2} + i\frac{\gamma_{1,2}+\gamma_{c1,2}}{2}$. To simplify our analysis, we choose $\omega_1 = \omega_2$ in the following discussion.

This simple two resonance/zero TCMT model can explain all of our results; it is new in the present work and was not part of the analysis used in ref. (*37*). Without tuning of parameters, this model will lead to two distinct pairs of eigenfrequencies for both the resonances and for the zeros, and no EPs. Generically, at any complex frequency, $\omega$, two complex eigenvalues $\lambda_{1,2}$ can be found for $S$ and the corresponding two component eigenvectors $v_{1,2}$ (with proper amplitudes and phases satisfying $Sv_{1,2} = \lambda_{1,2}v_{1,2}$) define the two eigenchannels, which are the specific superpositions of physical channels for which the waveform remains invariant during the scattering processes. Generically, at a resonance of S one eigenvalue of S diverges, and at a zero of S one eigenvalue vanishes; in each case the other eigenvalue is finite; however, at resonance each *element* of the S-matrix diverges, which, as noted, makes experimental study complicated. In Figs. 1C-G we illustrate the different kinds of EPs that can arise via parameter tuning in this model. In Fig. 1C we show the familiar case of lossless (no absorption) cavities with the coupling $\kappa = \left|\frac{\gamma_{c1}-\gamma_{c2}}{4}\right|$; in this case both the poles and zeros are simultaneously degenerate, i.e., at an EP, and the existence of the EP zero is not usually noted. Once absorption is added, the poles and zero frequencies are *not* complex conjugates and the EP condition for each generically differs. As the coupling, $\kappa$, is varied, the condition for a resonant EP is $\kappa = \left|\frac{(\gamma_{c1}-\gamma_{c2})+(\gamma_1-\gamma_2)}{4}\right|$ (supplementary text S2), and the condition for an EP zero is $\kappa = \left|\frac{(\gamma_{c1}-\gamma_{c2})-(\gamma_1-\gamma_2)}{4}\right|$ (supplementary text S3). This implies that, as the coupling varies, the EP zero will occur before the resonant EP if the differential coupling has the same sign as the differential absorption between the two cavities, and vice versa if they have opposite sign. In Figs. 1D and 1E, we show schematically the former case; before the coupling is turned on the two zeros and poles all have the same real part of the frequency. As the coupling increases, first the two zeros meet at an EP (Fig. 1D), before the two poles meet; then the two zeros move apart horizontally, while the two poles merge at a higher value of the coupling (Fig. 1E). This scenario corresponds to the data shown in Figs. 2A and 2B, to be discussed further below. Because the two EPs don't coincide parametrically and the absorbing EP is away from the real axis, the absorption lineshape shows a characteristic peak instead of a dip near zero detuning, not seen in previous experiments, as shown in Figs. 2C and 2D. The case of special interest is when the two zeros meet on the real axis, corresponding to steady-state perfect absorption at an EP, i.e., CPA EP (Fig. 1F). This happens, within TCMT, at the critical coupling condition, when the total intrinsic loss $\gamma_1 + \gamma_2$ is balanced by the total radiative coupling $\gamma_{c1} + \gamma_{c2}$, which corresponds to an intracavity coupling value, $\kappa = \left|\frac{\gamma_{c1}-\gamma_1}{2}\right|$ (supplementary text S4). In general this leads to an anomalous quartic absorption lineshape (and in other geometries, to chiral absorption) (*37*). This is the situation we refer to as a generic CPA EP (Fig. 1F), following the terminology of (*37*).

However, for symmetric coupling rates ($\gamma_{c1} = \gamma_{c2}$), one sees from the conditions given above, that, within the TCMT model, one finds (Fig. 1G, supplementary text S5) that both the zeros and poles coalescence at the *same value* of $\kappa$, (although the pole is off the real axis) and we do have simultaneous resonant and absorbing EPs. Nonetheless, the absorption properties are dominant in scattering, due to the EP zero being at a real frequency. This case also coincides with an EP of S, in which the eigenvectors of the S matrix and not just the eigenfunctions of the wave operator coalesce (*37*) (supplementary text S5). We refer to this as non-generic CPA EP. The singular behavior of the S-matrix leads to a distinct scattering behavior from the generic case to be discussed and demonstrated experimentally below. This case can be understood in terms of gauged PT-symmetry (*39*), where we expect all three types of EPs to coincide (*5, 21, 29, 40*). The three cases illustrated in Figs. 1D-F above have not been realized in previous experiments.

**Resonant EPs and absorbing EPs**

Here we experimentally realize and compare the cases illustrated in Figs. 1D-F for the resonant EPs and the absorbing EPs in the setup shown in Fig. 1B; initially, for simplicity, we study the case with the bottom waveguide decoupled from $\mu R_2$, which implies, in the S-matrix terminology, that only reflection or absorption of the input is possible. The resonant frequencies are aligned by the temperature control of $\mu R_2$ via a thermo-electric cooler (TEC). The reflection coefficient, the only non-zero element of $S$, is measured as a function of laser frequency detuning in order to derive the coupling strength $\kappa$, the poles and the zeros. The resonant EP and absorbing EP can be distinguished by the phase transitions of the complex poles and zeros when the intercavity coupling strength $\kappa$ is varied (Fig. 2, A and B). The parameters here correspond to the scenario where the zeros meeting at a lower coupling, $\kappa = \kappa_{th}$, and then move apart; the coupling is increased further and then the poles meet at $\kappa \approx 1.32\kappa_{th}$. For each case, after the transition the zeros/poles no longer have the same real part of the frequency, but instead share approximately the same imaginary part. This clearly confirms that, in the coupled cavity system, resonant EPs and absorbing EPs generally occur at different coupling strengths (or equivalently, different differential loss rates). In the alternative scenario, mentioned above, the absorbing EP occurs after the appearance of resonant EP as $\kappa$ increases (supplementary Fig. S2).

It is worth noting that the single dip/peak lineshape of the output signal spectrum is often regarded as a signature of resonant EPs. Particularly, in a single cavity with coupled CW and CCW modes (*5*) or parity-time (PT) symmetric coupled microcavities (*6, 41*), the transmission spectrum evolves from a doublet to a single dip at EPs. However, for a general system of two coupled units with arbitrary loss (or gain), EPs may not be accompanied by a single dip/peak in the transmission/reflection spectra. This is confirmed experimentally by the spectra of the scattered signal at the absorbing EP (Fig. 2C) and the resonant EP (supplementary Fig. S3), which both exhibit an absorbing doublet. This is attributed to the fact that the zeros and poles are not complex conjugate pairs and do not coalesce simultaneously, so that the minimum output is not reached at the frequencies $Re(\omega_{z1,2})$ or $Re(\omega_{p1,2})$ (see supplementary text S7). This observation can also be explained by studying the simulated field redistribution versus the frequency detuning at the absorbing EP (Fig. 2D). At zero detuning, the optical field tends to localize in $\mu R_2$ (which is true for resonant EPs as well, since the field distribution is not only influenced by the eigenstate of the Hamiltonian, but also by how the system is probed (*42*)). Away from the zero detuning, the intracavity energy $E_2$ reduces sharply due to the narrow linewidth of $\mu R_2$, while $\mu R_1$, being the only source to couple energy into $E_2$, restores more energy in itself. The total dissipation thereby increases because the redistributed field experiences larger loss in $\mu R_1$ ($\gamma_1 > \gamma_2$). Therefore, we find a reduction of the output power (Fig. 2C) when the detuning deviates from zero. This behavior occurs because neither EP occurs on the real axis, and will disappear for the next case, CPA EP.

**One-channel and two-channel CPA EPs**

We also study CPA EPs, where, as discussed above, critical coupling is achieved, and the two degenerate zeros occur at a real frequency. With the scheme in Fig. 3A, we realize a one-channel generic CPA EP (EP critical coupling) by changing the $\mu R_1$-waveguide gap to reach the CPA condition ($\gamma_{c1} = 119.2814 MHz \approx \gamma_1 + \gamma_2$), and adjusting the intercavity gap to achieve an absorbing EP ($\kappa = \left|\frac{\gamma_{c1}+\gamma_2-\gamma_1}{4}\right|$), differing from the condition noted above due to $\gamma_{c2} = 0$ here). As predicted, the measured spectrum is well-fit by a quartic function, which verifies the anomalous broadband absorption lineshape for CPA EPs (Fig. 3B and supplementary text S8). In comparison, the Lorentzian curve fitting result based on the measured $\gamma_{c1}$ shows a large deviation from the data (Fig. 3B).

CPA is a more general phenomenon than critical coupling to a single channel as shown here. It is known that if the correct eigenchannel of S is incident, multichannel destructive wave interference will occur, causing absorption to be the only remaining loss channel. To study two-channel CPA EPs, we restore the coupling to the second waveguide (Fig. 3C). Now CPA EP will occur for the two-component eigenchannel, which requires simultaneous coherent illumination via ports 1 and 3. It is not straightforward in this setup to find the parameter values to realize the generic case of CPA EP, found for the one-channel case above. However, we can find a signature for the non-generic CPA EP ($\gamma_{c1} = \gamma_{c2}$), because, as noted, it occurs simultaneously with the resonant EP. The signature of a resonant EP is the merging of a doublet into a single peak in the spectrum of $t_1$ ($t_2$) when tuning $\kappa$, under single port illumination. As shown in supplementary text S6, $t_{1,2}$ are only determined by the position of the poles and not the zeros under this excitation condition. Thus, in the experiment, we first set symmetric waveguide-cavity coupling and fulfill the critical coupling condition, i.e., $\gamma_{c1} = \gamma_{c2} = (\gamma_1 + \gamma_2)/2$, and then tune $\kappa$ (by repositioning $\mu R_2$ in parallel to the waveguide) to find the resonant EP. The non-generic CPA EP is thereby realized. It is known that at a non-generic CPA EP, S takes the universal form with the four elements having the same amplitudes, i.e., $|r_1| = |t_1| = |t_2| = |r_2|$ (37). Experimentally we observe this effect by characterizing $|r_1|^2$ and $|t_1|^2$ with only port 1 excited, as well as $|r_2|^2$ and $|t_2|^2$ with only port 3 excited (Figs. 4A and 4B). $|r_1|^2$ and $|t_1|^2$ are found almost equal at the zero detuning, while $|r_2|^2$ and $|t_2|^2$ approach each other with a small discrepancy due to imperfections such as backscattering effects in the cavity modes. According to the theory, the perfect absorption of the input wave is achieved by coupling the input into the absorbing eigenchannel defined by the eigenstate of $S$, i.e., $[1, -i]$ (for $\gamma_1 < \gamma_2$), which represents balanced amplitudes and a $\pi/2$ phase difference for the two inputs. This eigenstate still holds for nonzero detuning (supplementary text S5). In the experiment, we balance the input power in the two channels using optical attenuators (see Materials and Methods in the supplementary material for details) and control the relative phase by an electro-optic phase modulator (EOM). The measured output spectra from port 2 and port 4 exhibit similar lineshape (Fig. 3D) due to symmetric coupling rates (supplementary text S8), which is a unique feature of non-generic CPA EPs. Strong absorption is observed; the non-ideal perfect absorption is due to the fluctuation of optical phases and coupling parameters. Moreover, the total absorption lineshape for the eigenchannel input is quadratic instead of quartic, because the eigenvalues of $S$ are not analytic around the zero detuning (supplementary text S8).

**Phase response at the two-channel CPA EP**

The degree of absorption in the case of two-channel CPA (at an EP or not) is sensitive to the relative phase of the inputs, and should be maximal when the phase difference corresponds to the relevant eigenchannel of S. To characterize the phase response of our system operating at a CPA EP, we measure the output power from port 2 and port 4 around zero detuning normalized to the total input power ($P_2$ and $P_4$), as the relative phase between the two inputs is varied (see details in Materials and Methods in the supplementary material). The two output signals oscillate in phase with the relative phase change, consistent with the theoretical prediction (Fig. 4C), and reach their minimum at $\phi = -\pi/2$. The two outputs add up to the total output ($P_{tot}$) with a large oscillation against the relative phase (Fig. 4C). Furthermore, the modulation depth, defined as the difference between the maximum and minimum output power with varying relative phase normalized to the total input power, is measured versus the detuning frequency. The modulation depth of the output from port 2 ($M_2$) is found to exhibit a single peak, while that from port 4 ($M_4$) shows a doublet (Fig. 4D). The difference in the lineshapes is owing to the fact that the input vectors that maximize and minimize the output power at a nonzero detuning are not always the eigenstate of $S$, and thus the feature of the non-generic CPA EP is hidden. Furthermore, the spectrum of the total output modulation depth ($P_{tot}$) displays a flattened window with the modulation depth higher than 0.8 in a frequency range above 500 MHz (comparable with $\gamma_c$), which indicates a broadband large phase sensitivity at the CPA EP.

**Discussion and outlook**

Our study suggests that in order to verify the presence of EPs and reveal their properties, it is extremely important to properly engineer the input. With a single-port input, reflection ($|r_{1,2}|^2$) and transmission ($|t_{1,2}|^2$) spectra can be measured to quantify the basic scattering properties. $|r_{1,2}|^2$ is influenced by both poles and zeros and hence cannot easily be used to infer the presence of resonant or absorbing EPs. Nevertheless, as noted, $|t_{1,2}|^2$ is only influenced by the poles (resonances) and thus can be employed to search for resonant EPs. Furthermore, more complicated scattering properties are revealed when the system is probed from multiple channels. With the eigenstate of the S matrix as the input, a generic CPA EP induces perfect absorption with a quartic-lineshape spectrum, whereas a non-generic CPA EP leads to quadratic lineshapes in the output spectra from different ports. The signature scattering properties for different EPs are critical for non-Hermitian applications such as EP enhanced sensing, where the amplified frequency splitting in the output spectrum can only be observed with a proper probing scheme.

The physics introduced in this study can be applied to other platforms including grating structures, metamaterial/metasurfaces, acoustic systems, electrical circuits, etc., where CPA EPs enable phase-sensitive single-mode absorption with quartic lineshapes. Furthermore, the realization of resonant, absorbing and CPA EPs in a single system offers rich physics and versatile means for the control of scattering behavior in various non-Hermitian wave systems. Moreover, the absorbing EPs can shed light on non-Hermitian systems in the regime where quantum fluctuations are important. It has been pointed out that realizing pure PT symmetry in quantum photonic systems by coupling gain and loss units is not possible, since introducing gain will unavoidably introduce additional noise that breaks the PT-symmetry (*43*). This hinders the exploration of resonant EPs via emission features in quantum systems. However, by engineering the quantum interference and dissipation, one can realize the absorbing EPs without the need for gain in a quantum system. The absorbing EPs in quantum photonic structures are expected to modify the absorption and scattering properties of nonclassical light, which can be utilized to engineer quantum channels for the suppression of quantum states or noises and enhance nonlinear light-matter interaction at the quantum level. Hence such novel EPs could open up new avenues for nonclassical light generation (*44*, *45*), quantum state control (*46*, *47*) and quantum storage (*48*). It would be of special interest to explore the effects of such EPs on quantum behavior including quantum interference (*49*), decoherence (*50*), and the evolution of entanglement (*51*), etc.

**Acknowledgement:** This work was supported by the NSF grant No. EFMA1641109. A.D.S. acknowledges the support of the CMMT grant DMR-1743235.

**Author contributions:** L.Y. and A.D.S conceived the joint project. C.W. and L.Y. designed the experiments with the help from W.R.S. and A.D.S. C.W. performed the experiments. C.W. analysed experimental data with help from W.R.S. Theoretical background and simulations were provided by C.W. and W.R.S. All authors discuss the results and wrote the manuscript. L.Y. supervised the project.

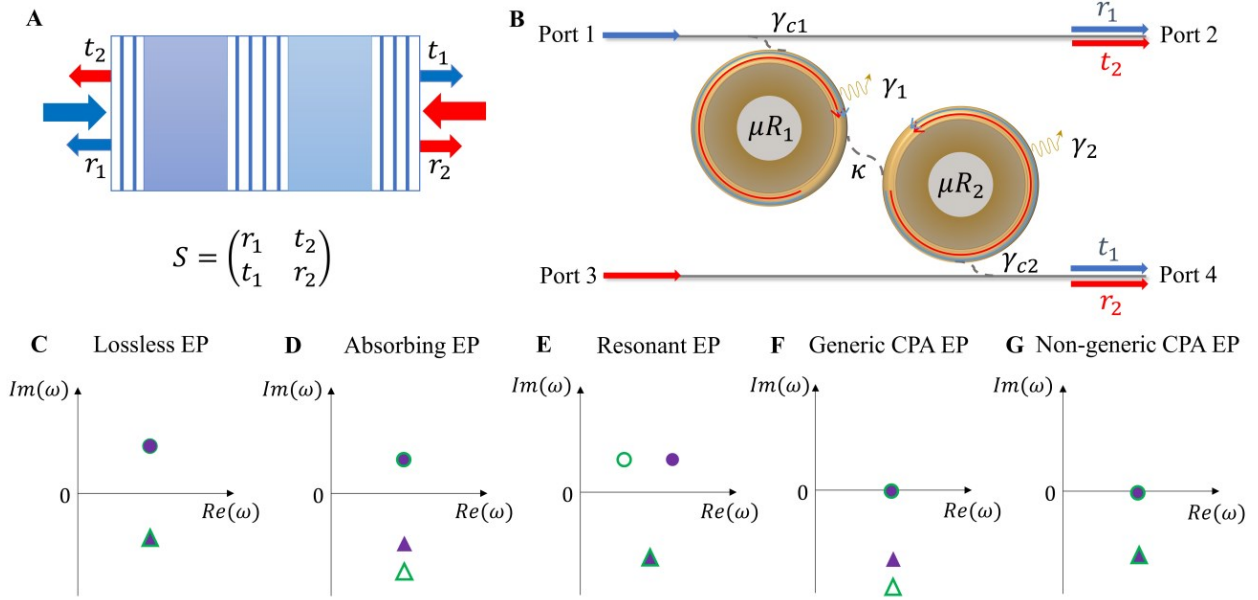

**Fig. 1. Scattering properties of the coupled microcavity systems.** (**A**) Scattering from two one-dimensional cavities formed by external Bragg mirrors coupled via a central, partially reflecting Bragg mirror, studied in ref. (*37*). The scattered output from the incident waves in the left and right channels determine the S matrix. (**B**) Schematic diagram of the analogous model to (A) realized with coupled WGM microcavities ($\mu R_1$ and $\mu R_2$), with double fiber taper waveguides as the input and output channels. The two cavities with intrinsic loss rates $\gamma_1$ and $\gamma_2$ are coupled to the waveguides with coupling strengths $\gamma_{c1}$ and $\gamma_{c2}$, respectively. $\kappa$ denotes the intercavity coupling strength. The analogues of the reflection and transmission amplitudes are defined in the diagram (see discussion in the text). (**C**)-(**G**) Zeros (circles) and poles (triangles) of the system's S matrix in the complex frequency plane for different types and configurations of EPs (see discussion in the text). (C) Without absorption or gain, poles and zeros appear in conjugate pairs and will coalesce at the same parameter values, at complex conjugate frequencies. (D), with differential absorption loss in the system, generically, zeros can coalesce at an absorbing EP at a lower intercavity coupling than the poles as shown. (E) At a larger coupling, the poles meet at a resonant EP, but the zeros are now distinct and have different real parts of their frequencies. (F) A generic CPA EP occurs when the critical coupling condition is satisfied and the zeros coalesce at a real frequency. (G) A non-generic CPA EP is realized when, in addition to critical coupling leading to a real absorbing EP, due to the external coupling rates being equal, the resonances are also at an EP (away from the real axis).

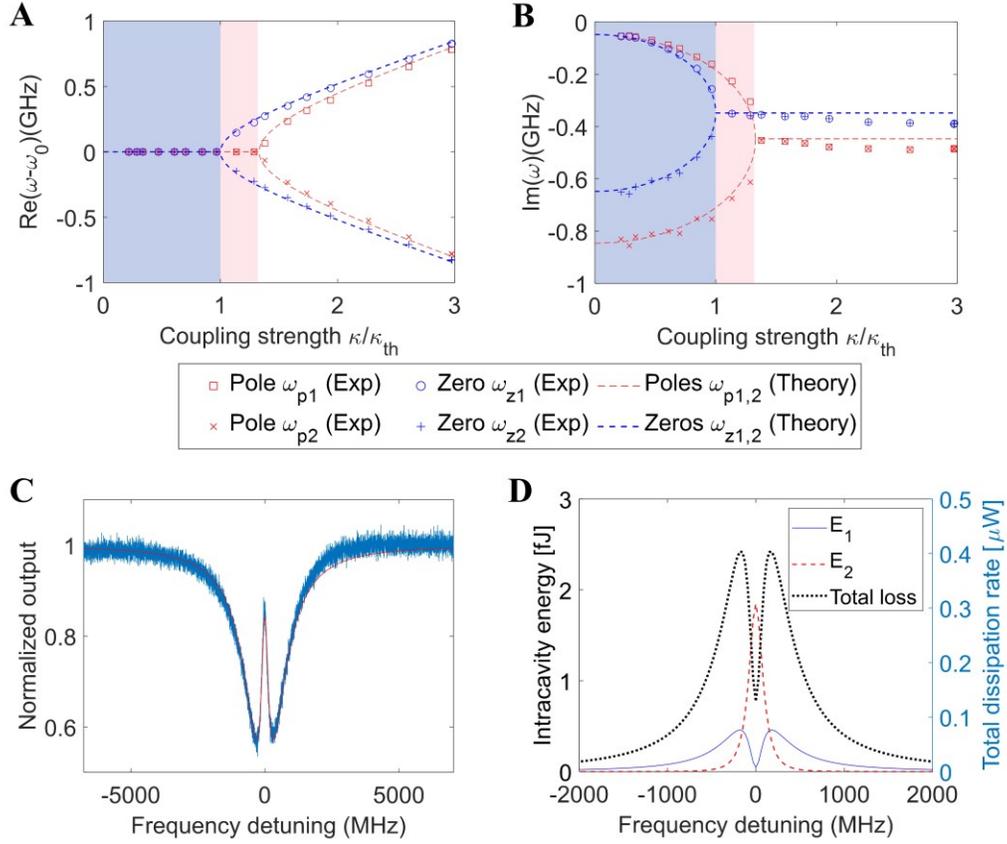

**Fig. 2**. **Characterization of resonant EPs and absorbing EPs.** (**A** and **B**) Experimentally and theoretically obtained phase transition diagrams for the real parts (A) and imaginary parts (B) of the poles and zeros as a function of normalized coupling strength $\kappa/\kappa_{th}$. "Exp"/"Theory" in the legend refer to experimental/theoretical results. Parameters: $\gamma_1 = 1494.1 MHz$, $\gamma_2 = 96.3 MHz$, $\gamma_{c1} = 193.6 MHz$, $\gamma_{c2} = 0$. (**C**) Normalized output spectrum near an absorbing EP. The blue and red curves are experimental and curve fitting results, respectively. (**D**) Simulation results of the energy distribution in $\mu R_1$ ($E_1$) and $\mu R_2$ ($E_2$), and the total dissipation rate, as a function of the frequency detuning at the absorbing EP. The input power is $1 \mu W$.

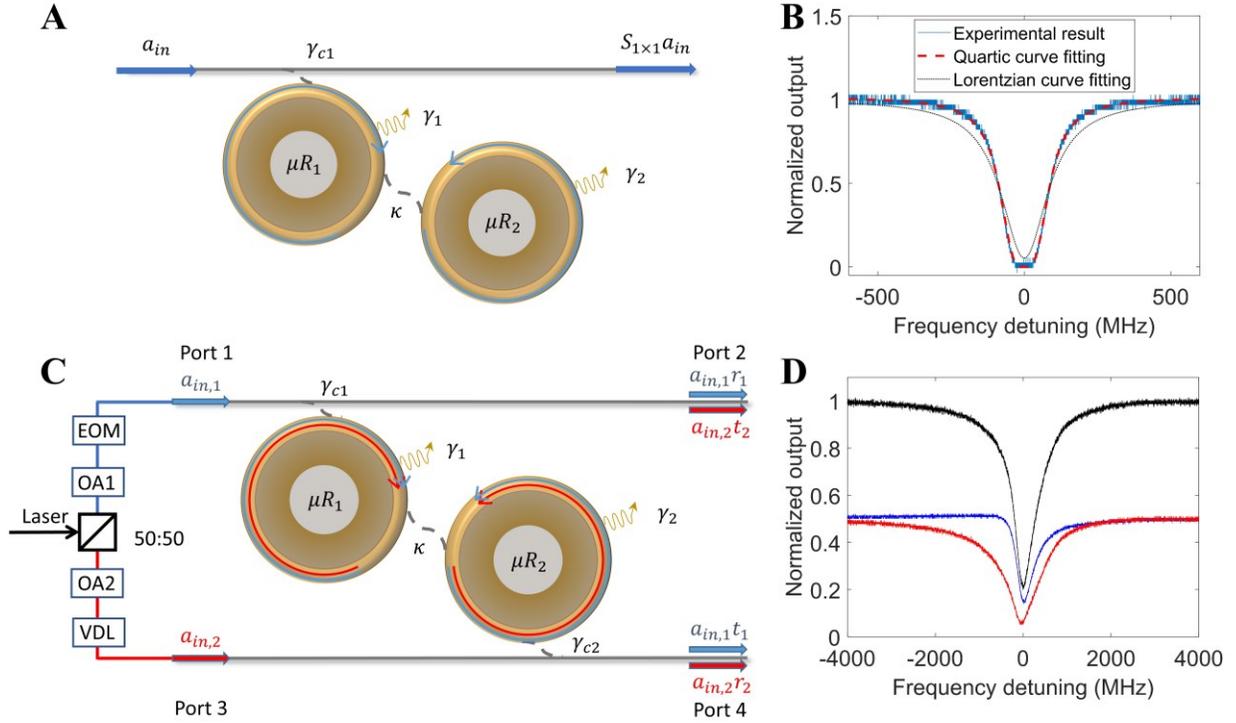

**Fig. 3. Coherent perfectly absorbing exceptional points (CPA EPs).** (**A**) Schematic diagram of the experimental setup for a one-channel CPA EP. The S matrix is reduced to only one element $S_{1\times 1}$. (**B**) Experimentally obtained normalized output spectrum at the one-channel CPA EP and the curve fitting results using quartic and Lorentzian functions. (**C**) Schematic diagram of the experimental setup for a two-channel CPA EP. The magnitudes of the two inputs are controlled by two variable optical attenuators (OA1 and OA2). The relative phase of the two inputs is controlled by an electro-optic phase modulator (EOM). The lengths of the two optical paths are balanced by a variable optical delay line (VDL). (**D**) Experimentally obtained spectra of the output from port 2 (blue curve), output from port 4 (red curve), and the total output (black curve) normalized to the total input power at the non-generic CPA EP. To excite the system by eigenvector of S, the amplitudes and relative phase of the two input fields are adjusted by OA1, OA2 and EOM.

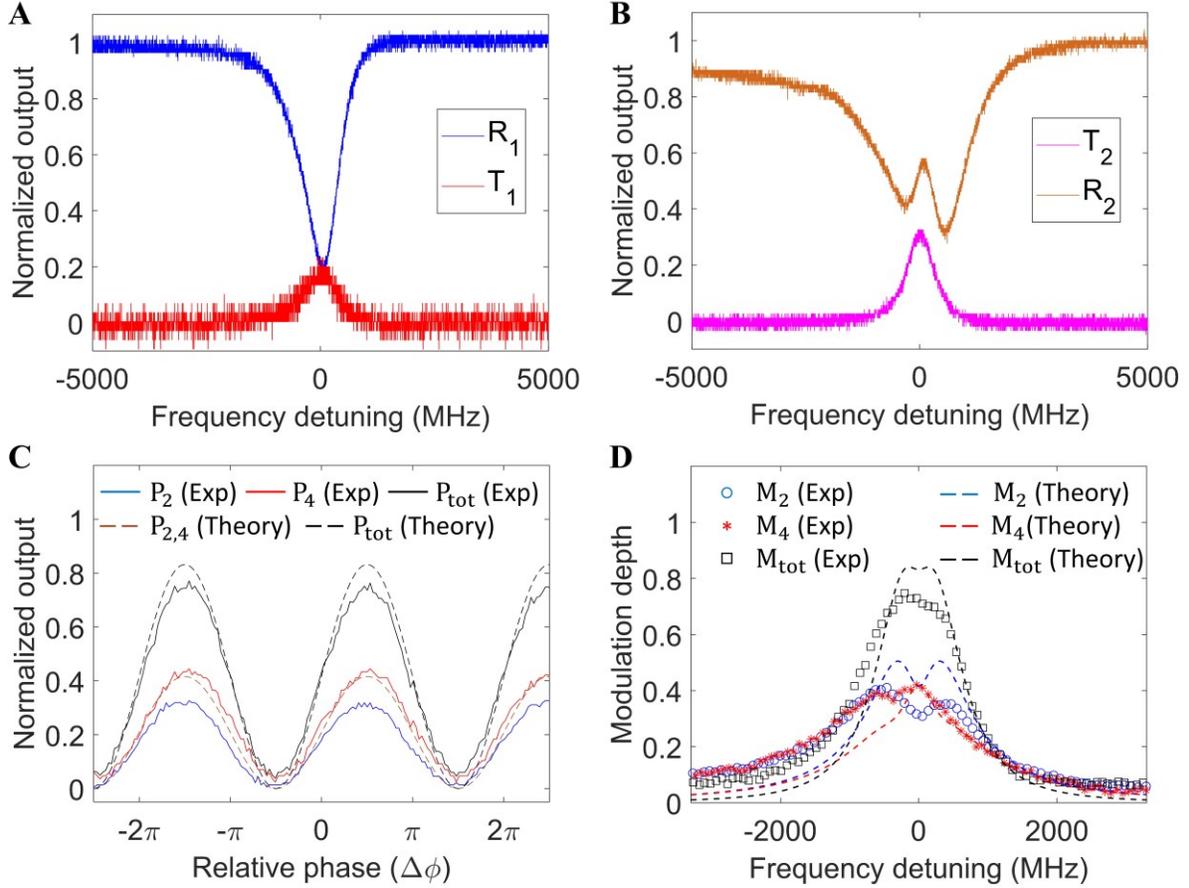

**Fig. 4. Scattering properties and phase modulation at the two-channel CPA EP.** (**A**) Experimentally measured spectra of $R_1 = |r_1|^2$ and $T_1 = |t_1|^2$ by exciting port 1 only. (**B**) Experimentally measured spectra of $R_2 = |r_2|^2$ and $T_2 = |t_2|^2$ by exciting port 3 only. (**C**) Experimentally obtained output power from port 2 ($P_2$) and port 4 ($P_4$) at the zero detuning ($\delta = 0$) normalized to the total input power as a function of the relative phase. $P_2$ and $P_4$ are equal in TCMT. The total normalized output power ($P_{tot}$) is obtained by summing up $P_2$ and $P_4$. (**D**) Modulation depth – the difference between the maximum and the minimum power normalized to the total input power under varying relative phase – of the output from port 2 ($M_2$) and the output from port 4 ($M_4$) versus frequency detuning. The modulation depth of the total output power ($M_{tot}$) is not the sum of $M_2$ and $M_4$ because at each frequency detuning, the relative phases which maximize (or minimize) the two outputs can be different.

## Supplementary Materials

**Materials and Methods**

Balancing optical power in the two input channels

The CPA operation is sensitive to both the amplitudes and phases of the input signals (*1–10*). In the case of a non-generic CPA EP, the eigenvector of the S matrix is given by

$$v = \begin{pmatrix} \pm i \\ 1 \end{pmatrix}, \quad (S1)$$

where we take the plus sign if $\gamma_1 < \gamma_{c1}$, and the minus sign otherwise (see supplementary text S8). As a result, the optical fields injected from port 1 and port 3 should have equal amplitudes and a phase difference of $\pm \pi/2$.

In our experimental setup, since silica fiber tapers are used as the input and output channels, the optical loss is not negligible along the guided wave propagation. The losses in the two input channels could be unequal due to the unbalanced optical path lengths or different loss in the tapered fibers. Thus, the optical power at the two taper-cavity coupling points could be different. To overcome the loss imbalance and achieve equal input amplitudes, we utilize thermo-optic effects as a reference of the input power at the taper-cavity coupling points.

Based on the setup shown in Fig. S1, we remove $\mu R_2$ and use $\mu R_1$ only to calibrate the loss difference in the two optical paths. We first tune the variable optical attenuator in the first optical path (OA1) to adjust the optical input power to the level of $100 \mu W$ so that the cavity enters the nonlinear regime with strong thermal effects. We critically couple $\mu R_1$ to the first fiber taper, and a thermal triangle lineshape appears in the transmission spectrum when the laser is scanned from shorter to larger wavelengths due to the thermo-optic effect (*11–15*). Under the fixed cavity-taper coupling condition, the width of the triangle has a deterministic relation with the input power. Then we decouple the $\mu R_1$ from the first taper and critically couple $\mu R_1$ to the second taper, where we can also observe a thermal triangle in the transmission spectrum. We then tune the variable optical attenuator in the second optical path (OA2) so that the triangle has the same width as measured in the first one. After that, we use a photodetector to measure the output power coming out from OA1 and OA2, respectively. Assuming linear optical loss in the optical paths, the ratio between the two output powers will remain invariant with arbitrary input power from the laser. During the CPA EP measurement, we maintain the ratio between the optical power coming out of the two attenuators and reduce the laser power dramatically to the $\mu W$ level so that the cavity enters the linear regime with a negligible influence of thermo-optic effects. Thereby, the balance of the input field amplitudes is achieved.

Measurement of scattered output signals versus the relative phase

In the measurement of CPA EP, the balance between the optical path lengths is critical for obtaining a clear spectrum of the output signal. To see this, we investigate the case where the two optical path lengths are different by $\Delta l$, and the laser frequency is scanned at the rate of $v$ [Hz/s]. The frequency difference between the two beams arriving at the two taper-cavity coupling points is given by

$$\Delta f = v \frac{n \Delta l}{c}, \quad (S2)$$

where $n$ is the average refractive index of the media in the optical paths. The different frequencies for the light in the cavities will lead to a beat note in the output signals, with a period of $\frac{2}{\Delta f}$. Consequently, the relative phase between the two input signals will also vary as $\Delta\phi = 2\pi\Delta f t + \phi_0$, which leads to an oscillation in the spectrum of the scattered output signals. In the measurement for Fig. 3D in the main text, we need to keep $\Delta l = 0$ by tuning the variable optical delay line (VDL) (Fig. S1) so that the relative optical phase between the two optical paths is fixed as the laser frequency is scanned.

However, the oscillation in the spectrum induced by unbalanced optical paths could be useful for investigating the system's response to the relative phase, as shown in Fig. 4 in the main text. If $\Delta l$ is sufficiently large, the oscillation period is much smaller than the time it takes to scan a linewidth of the cavity mode ($\Delta v$), i.e.,

$$\frac{2}{\Delta f} \ll \frac{\Delta v}{\dot{v}}, \tag{S3}$$

which yields

$$\Delta l \gg \frac{2c}{n\Delta v}. \tag{S4}$$

Then within a small range of frequency scanning, the relative phase could be modulated over multiple periods. Thus, the detuning is approximately constant within one period of relative phase oscillation. The response of the scattered signal to the relative phase can then be measured at any frequency within the scanning range. In our experiment, $\dot{v} \approx 2400\text{GHz/s}$, $n \approx 1.45$, $\Delta v > 40\text{MHz}$, therefore the condition works well if $\Delta l \gg 10m$, which can be fulfilled by inserting long fibers into one optical path as delay lines. Meanwhile, to avoid the effect of the frequency difference between the two input fields, we need to make sure that $\Delta f \ll \Delta v$, i.e., $\Delta l \ll \frac{c\Delta v}{n\dot{v}} = 3448.3m$, which is not hard to satisfy in the experiments.

In the measurement for Fig. 4C in the main text, we adopt this method, that is, to intentionally break the balance of the lengths of the two optical channels by inserting fibers as delay lines (Fig. S1), and to measure the output power spectra from port 2 and 4 within a small frequency range near the zero detuning, which shows a response as a function of the relative phase between the two inputs. The result can be considered as the signal at zero detuning against the relative phase, to a good approximation. The data can also be used to derive Fig. 4D in the main text, by extracting the maximum and minimum output power at different frequency detuning.

In principle, an alternative method is to apply a sinusoidal phase modulation via EOM and record the spectra of the output signals at each relative phase, followed by extracting the maximum and minimum at each detuning. However, several factors may cause inaccuracy to this method. First, in order to quantify the exact phase value at each frame, one need to precisely set the voltage amplitude of EOM to be $V_{pi}$, which is the voltage to generate a $\pi$ phase shift. Second, due to the finite number of frames of data collection and limited measurement speed, the maximum and minimum output power cannot be precisely captured.

**Supplementary Text**
Theoretical analysis of various kinds of exceptional points (EPs) in the coupled microcavity system

## S1. Theoretical framework

We build a theoretical framework based on the temporal coupled-mode theory (TCMT) (*16*) to study the scattering behavior of the coupled microcavity system. The Hamiltonian describing the two directly coupled microcavities with resonant frequencies $\omega_{1,2}$ and intrinsic loss rates $\gamma_{1,2}$ can be written as

$$H_0 = \begin{pmatrix} \omega_1 - i\frac{\gamma_1}{2} & \kappa \\ \kappa & \omega_2 - i\frac{\gamma_2}{2} \end{pmatrix}, \quad (S5)$$

where $\kappa$ is the coupling strength between the two microcavities. The coupling channels to the waveguides can be described by a diagonal coupling matrix

$$D = diag(\sqrt{\gamma_{c1}}, \sqrt{\gamma_{c2}}). \quad (S6)$$

Based on the TCMT, the scattering (S) matrix is then given by

$$S = 1 - iD^\dagger \frac{1}{\omega - \left[H_0 - \frac{iD^\dagger D}{2}\right]} D, \quad (S7)$$

where the term $H_0 - iD^\dagger D/2$ gives the definition of the effective Hamiltonian that involves the optical dissipation into the coupling channels

$$H_{eff} = H_0 - \frac{iD^\dagger D}{2} = \begin{pmatrix} \omega_1 - i\frac{\gamma_1 + \gamma_{c1}}{2} & \kappa \\ \kappa & \omega_2 - i\frac{\gamma_2 + \gamma_{c2}}{2} \end{pmatrix}. \quad (S8)$$

In general, we can derive the poles and zeros of $S$ as well as the conditions of resonant and absorbing EPs based on Eq. S8. We note that for either type of EPs, multiple solutions can be obtained based on the conditions for the parameters. For simplicity, we consider a special case that the two cavity resonances match with each other, i.e., $\omega_1 = \omega_2$. Other cases which break this restriction may be of interest for studying EPs in other schemes, for example anti-PT symmetry (*17–26*).

## S2. Resonant EP

We first calculate the eigenvalues of $H_{eff}$, which governs the resonances of the system,

$$\lambda_{r\pm} = \frac{\omega_1 + \omega_2}{2} - i\frac{\gamma_1 + \gamma_{c1} + \gamma_2 + \gamma_{c2}}{4}$$

$$\pm \frac{1}{2}\sqrt{\left(\omega_1 + \omega_2 - i\frac{\gamma_1 + \gamma_{c1} + \gamma_2 + \gamma_{c2}}{2}\right)^2 - 4\left(\omega_1 - i\frac{\gamma_1 + \gamma_{c1}}{2}\right)\left(\omega_2 - i\frac{\gamma_2 + \gamma_{c2}}{2}\right) + 4\kappa^2}$$

$$= \frac{\omega_1 + \omega_2}{2} - i\frac{\gamma_1 + \gamma_{c1} + \gamma_2 + \gamma_{c2}}{4} \pm \frac{1}{2}\sqrt{\left(\omega_1 - \omega_2 - i\frac{\gamma_1 + \gamma_{c1} - \gamma_2 - \gamma_{c2}}{2}\right)^2 + 4\kappa^2}, \quad (S9)$$

which are also the poles of $S$. When $\omega_1 = \omega_2 = \omega_0$, the resonances are given by

$$\lambda_{r1,2} = \omega_0 - i\frac{\gamma_1 + \gamma_{c1} + \gamma_2 + \gamma_{c2}}{4} \pm \sqrt{\kappa^2 - \left(\frac{\gamma_1 + \gamma_{c1} - \gamma_2 - \gamma_{c2}}{4}\right)^2}, \qquad (S10)$$

which are equal to the poles of $S$ ($\omega_{p1,2}$). Then we derive the conditions for the resonant EPs

$$\kappa = \frac{|\gamma_1 + \gamma_{c1} - \gamma_2 - \gamma_{c2}|}{4}. \qquad (S11)$$

### S3. Absorbing EP

Now we turn to the calculation of zeros of $S$. For simplicity, we define $\Omega_{1,2} = \omega_{1,2} - i\frac{\gamma_{1,2} + \gamma_{c1,2}}{2}$. It follows that

$$\frac{1}{\omega - H_{eff}} = \begin{pmatrix} \omega - \Omega_1 & -\kappa \\ -\kappa & \omega - \Omega_2 \end{pmatrix}^{-1} = \frac{1}{\det(\omega - H_{eff})}\begin{pmatrix} \omega - \Omega_2 & \kappa \\ \kappa & \omega - \Omega_1 \end{pmatrix}. \qquad (S12)$$

Thus, we can obtain the S-matrix

$$S = 1 - i\begin{pmatrix} \sqrt{\gamma_{c1}} & 0 \\ 0 & \sqrt{\gamma_{c2}} \end{pmatrix}\begin{pmatrix} \Delta_2 & \kappa \\ \kappa & \Delta_1 \end{pmatrix}\begin{pmatrix} \sqrt{\gamma_{c1}} & 0 \\ 0 & \sqrt{\gamma_{c2}} \end{pmatrix}\frac{1}{\Delta_1\Delta_2 - \kappa^2}$$

$$= \begin{pmatrix} 1 - i\frac{\gamma_{c1}\Delta_2}{\Delta_1\Delta_2 - \kappa^2} & -i\frac{\sqrt{\gamma_{c1}\gamma_{c2}}\kappa}{\Delta_1\Delta_2 - \kappa^2} \\ -i\frac{\sqrt{\gamma_{c1}\gamma_{c2}}\kappa}{\Delta_1\Delta_2 - \kappa^2} & 1 - i\frac{\gamma_{c2}\Delta_1}{\Delta_1\Delta_2 - \kappa^2} \end{pmatrix}. \qquad (S13)$$

where $\Delta_{1,2} = \omega - \Omega_{1,2}$. The S-matrix is symmetric in our study which indicates the reciprocal wave transport. We can then calculate the eigenvalues of $S$

$$\sigma_{1,2} = 1 - i\frac{\gamma_{c1}\left(\delta_2 + i\frac{\gamma_2 + \gamma_{c2}}{2}\right) + \gamma_{c2}\left(\delta_1 + i\frac{\gamma_1 + \gamma_{c1}}{2}\right)}{2(\Delta_1\Delta_2 - \kappa^2)}$$

$$\pm \frac{1}{2(\Delta_1\Delta_2 - \kappa^2)}\sqrt{-\left(\gamma_{c1}\left(\delta_2 + i\frac{\gamma_2 + \gamma_{c2}}{2}\right) - \gamma_{c2}\left(\delta_1 + i\frac{\gamma_1 + \gamma_{c1}}{2}\right)\right)^2 - 4\gamma_{c1}\gamma_{c2}\kappa^2}. \qquad (S14)$$

If $\omega_1 = \omega_2 = \omega_0$, the eigenvalues of $S$ are given by

$$\sigma_{1,2} = \frac{\left(\left(\delta + i\frac{\gamma_1}{2}\right)\left(\delta + i\frac{\gamma_2}{2}\right) + \left(\frac{\gamma_{c1}}{2}\right)\left(\frac{\gamma_{c2}}{2}\right) - \kappa^2\right)}{(\Delta_1\Delta_2 - \kappa^2)}$$

$$\pm \frac{1}{(\Delta_1\Delta_2 - \kappa^2)}\sqrt{-\left(\frac{\gamma_{c1}}{2}\left(\delta + i\frac{\gamma_2 + \gamma_{c2}}{2}\right) - \frac{\gamma_{c2}}{2}\left(\delta + i\frac{\gamma_1 + \gamma_{c1}}{2}\right)\right)^2 - \gamma_{c1}\gamma_{c2}\kappa^2}, \qquad (S15)$$

where $\delta = \omega - \omega_0$.

For zeros, we have $\sigma_{1,2} = 0$, that is

$$\left(\left(\delta+i\frac{\gamma_1}{2}\right)\left(\delta+i\frac{\gamma_2}{2}\right)+\left(\frac{\gamma_{c1}}{2}\right)\left(\frac{\gamma_{c2}}{2}\right)-\kappa^2\right)$$

$$\pm\sqrt{-\left(\frac{\gamma_{c1}}{2}\left(\delta+i\frac{\gamma_2}{2}\right)-\frac{\gamma_{c2}}{2}\left(\delta+i\frac{\gamma_1}{2}\right)\right)^2-\gamma_{c1}\gamma_{c2}\kappa^2}=0, \qquad (S16)$$

which yields two solutions of zeros

$$\omega_{z1,z2}=\frac{\Omega_1+\Omega_2+i\gamma_{c1}+i\gamma_{c2}}{2}\pm\frac{1}{2}\sqrt{(\Omega_1+i\gamma_{c1}-\Omega_2-i\gamma_{c2})^2+4\kappa^2}$$

$$=\frac{\omega_1+\omega_2}{2}+i\frac{\gamma_{c1}+\gamma_{c2}-\gamma_1-\gamma_2}{4}\pm\frac{1}{2}\sqrt{\left(\omega_1+i\frac{\gamma_{c1}-\gamma_1}{2}-\omega_2-i\frac{\gamma_{c2}-\gamma_2}{2}\right)^2+4\kappa^2}. \qquad (S17)$$

The EP of zeros requires that

$$(\Omega_1+i\gamma_{c1}-\Omega_2-i\gamma_{c2})^2+4\kappa^2=0, \qquad (S18)$$

which follows that

$$\left(2\kappa+i(\Omega_1+i\gamma_{c1}-\Omega_2-i\gamma_{c2})\right)\left(2\kappa-i(\Omega_1+i\gamma_{c1}-\Omega_2-i\gamma_{c2})\right)=0. \qquad (S19)$$

Thus, two solutions are derived

$$\kappa_1=\frac{i(\omega_1-\omega_2)}{2}+\frac{\gamma_1+\gamma_{c2}-\gamma_2-\gamma_{c1}}{4}, \qquad (S20)$$

or

$$\kappa_2=\frac{i(\omega_2-\omega_1)}{2}+\frac{\gamma_{c1}+\gamma_2-\gamma_1-\gamma_{c2}}{4}. \qquad (S21)$$

For $\omega_1=\omega_2$, and $\kappa>0$, we have

$$\kappa=\frac{|\gamma_1+\gamma_{c2}-\gamma_2-\gamma_{c1}|}{4}. \qquad (S22)$$

It is worth noting that the critical value of $\kappa$ for an absorbing EP to occur, which is given by Eq. (S22), can be larger, smaller or identical compared to that for a resonant EP to occur, which is given by Eq. (S11). In the phase transition when we gradually increase $\kappa$, the absorbing EP is reached before the resonant EP if $\gamma_1>\gamma_2$, as we observe in Fig. 2 of the main text; but the resonant EP is reached before the absorbing EP if $\gamma_1<\gamma_2$, as shown by the experimental and simulation results presented in Fig. S2.

**S4. CPA EP**

Besides the conditions for the absorbing EPs, to further satisfy the requirement of a CPA EP, i.e., the absorbing EP occurs for real frequency, we need

$$0=Im\frac{(\Omega_1+\Omega_2+i\gamma_{c1}+i\gamma_{c2})}{2}=\frac{\gamma_{c1}+\gamma_{c2}-\gamma_1-\gamma_2}{4}, \qquad (S23)$$

which yields

$$\gamma_{c1}+\gamma_{c2}=\gamma_1+\gamma_2. \qquad (S24)$$

Therefore, we have

$$\kappa = \frac{|\gamma_1 - \gamma_{c1}|}{2}. \quad (S25)$$

We can finally summarize the conditions for CPA EP at $\omega_1 = \omega_2$

$$\gamma_{c1} + \gamma_{c2} = \gamma_1 + \gamma_2, \quad (S26a)$$

$$\kappa = \frac{|\gamma_1 - \gamma_{c1}|}{2}. \quad (S26b)$$

Under these conditions, we can evaluate the S matrix

$$S = A \begin{pmatrix} S_{11} & \pm i\sqrt{\gamma_{c1}\gamma_{c2}} \\ \pm i\sqrt{\gamma_{c1}\gamma_{c2}} & S_{22} \end{pmatrix}, \quad (S27)$$

where $A = \frac{1}{\left(\delta + i\frac{\gamma_1 + \gamma_{c1}}{2}\right)\left(\delta + i\frac{\gamma_2 + \gamma_{c2}}{2}\right) - \kappa^2} \frac{\gamma_1 - \gamma_{c1}}{2}$, $S_{11} = \left(\frac{\delta}{\frac{\gamma_1 - \gamma_{c1}}{2}} - i\right)\left(\delta - i\frac{\gamma_2 + \gamma_{c2}}{2}\right) - \frac{\gamma_1 - \gamma_{c1}}{2}$, and $S_{22} = \left(\delta - i\frac{\gamma_1 + \gamma_{c1}}{2}\right)\left(\frac{\delta}{\frac{\gamma_1 - \gamma_{c1}}{2}} + i\right) - \frac{\gamma_1 - \gamma_{c1}}{2}$. We take the positive sign if $\gamma_1 < \gamma_{c1}$, and the negative sign otherwise.

At zero detuning, $S$ becomes

$$S = A \begin{pmatrix} -\gamma_{c2} & \pm i\sqrt{\gamma_{c1}\gamma_{c2}} \\ \pm i\sqrt{\gamma_{c1}\gamma_{c2}} & \gamma_{c1} \end{pmatrix}, \quad (S28)$$

where A becomes a constant. The eigenvector of $S$ associated with the eigenvalue 0 is given by $v_1 = \begin{pmatrix} \pm i\sqrt{\gamma_{c1}} \\ \sqrt{\gamma_{c2}} \end{pmatrix}$; the eigenvector of $S$ associated with the eigenvalue $A(\gamma_{c1} - \gamma_{c2})$ is given by $v_2 = \begin{pmatrix} \pm i\sqrt{\gamma_{c2}} \\ \sqrt{\gamma_{c1}} \end{pmatrix}$. We take a positive sign if $\gamma_1 < \gamma_{c1}$, and a negative sign otherwise. One sees from this result that for a generic CPA EP ($\gamma_{c1} \neq \gamma_{c2}$) the input amplitudes from the two input ports are not balanced (unequal power), whereas for the non-generic case ($\gamma_{c1} = \gamma_{c2}$), they are balanced, making it easier to find the non-generic CPA EP input eigenvector, as we do in the experiment. Further, for the generic case the S-matrix has two distinct eigenvectors and is not at an EP, even though the wave-operator for incoming boundary conditions is at an EP. However, when $\gamma_{c1} = \gamma_{c2}$, one sees that the two eigenvalues each becomes zero and the two eigenvectors coalesce, implying that balanced input coupling is a sufficient condition to have an EP *of the scattering matrix* occur simultaneously with an EP of the wave operator. We note that this coincidence of the two kind of EPs is only found within the TCMT approximation and will not hold exactly when the exact S-matrix of a physical system is calculated due to the presence of other resonances (*27*).

### S5. Scattering EP

A Scattering EP happens at certain points in the parameter space and at certain frequencies, where the S-matrix becomes defective, the two eigenvalues are degenerate, and the two eigenvectors coalesce. The eigenvalue at the degeneracy is not in general zero. The general condition for this is:

$$\left(\frac{\gamma_{c1}}{2}\left(\delta + i\frac{\gamma_2}{2}\right) - \frac{\gamma_{c2}}{2}\left(\delta + i\frac{\gamma_1}{2}\right)\right)^2 + \gamma_{c1}\gamma_{c2}\kappa^2 = 0, \tag{S29}$$

which leads to

$$\delta\left(\frac{\gamma_{c1}}{2} - \frac{\gamma_{c2}}{2}\right) + i\frac{\gamma_2\gamma_{c1} - \gamma_1\gamma_{c2}}{4} = \pm i\sqrt{\gamma_{c1}\gamma_{c2}}\kappa. \tag{S30}$$

The condition for a scattering EP within this TMCT model is thus

$$\delta(\gamma_{c1} - \gamma_{c2}) = 0, \tag{S31}$$

and

$$\kappa = \pm \frac{\gamma_2\gamma_{c1} - \gamma_1\gamma_{c2}}{4\sqrt{\gamma_{c1}\gamma_{c2}}}. \tag{S32}$$

To prove consistency with our above results, we put in the condition for CPA EPs. Plugging Eq. (S24) into Eq. (S32), we have

$$\kappa = \pm \frac{(\gamma_{c1} + \gamma_{c2})(\gamma_{c1} - \gamma_1)}{4\sqrt{\gamma_{c1}\gamma_{c2}}}. \tag{S33}$$

Combining Eq. (S33) with Eq. (S25), we get

$$\pm \frac{(\gamma_{c1} + \gamma_{c2})(\gamma_{c1} - \gamma_1)}{4\sqrt{\gamma_{c1}\gamma_{c2}}} = \frac{|\gamma_1 - \gamma_{c1}|}{2}. \tag{S34}$$

To find a solution, we can only choose the sign to make the left-hand side positive. Then we have

$$\gamma_{c1} + \gamma_{c2} = 2\sqrt{\gamma_{c1}\gamma_{c2}}, \tag{S35}$$

which yields

$$\gamma_{c1} = \gamma_{c2}. \tag{S36}$$

As argued above, this result shows that for the CPA EP case, the $S$ matrix can be reduced to a defective form (within TCMT approximation) only if we have symmetric coupling strengths in the two channels. For $\gamma_{c1} = \gamma_{c2} = \gamma_c$, the $S$ matrix takes a defective form

$$S = A\begin{pmatrix} \frac{\delta^2}{\frac{\gamma_1 - \gamma_c}{2}} - \frac{i\gamma_c\delta}{\frac{\gamma_1 - \gamma_c}{2}} - \gamma_c & \pm i\gamma_c \\ \pm i\gamma_c & \frac{\delta^2}{\frac{\gamma_1 - \gamma_c}{2}} - \frac{i\gamma_c\delta}{\frac{\gamma_1 - \gamma_c}{2}} + \gamma_c \end{pmatrix}, \tag{S37}$$

where we take the positive sign if $\gamma_1 < \gamma_{c1}$, and the minus sign otherwise. The eigenvalue of $S$ is $2A\frac{\delta^2 - i\gamma_c\delta}{\gamma_1 - \gamma_c}$, and the corresponding eigenvector is

$$v = \sqrt{\gamma_c}\begin{pmatrix} \mp i \\ 1 \end{pmatrix}. \tag{S38}$$

The eigenvector is always the same for arbitrary detuning at the non-generic CPA EP; hence within TCMT we remain at an EP (just not CPA EP, with zero eigenvalue) as the frequency is varied near the CPA EP frequency. This prediction is not valid outside of the TCMT approximation, just as the S-matrix EP and the CPA EP do not exactly coincide in an exact scattering calculation (*27*). The symmetric coupling is also a necessary condition for an absorbing EP and a resonant EP to occur at the same time, again within the TCMT approximation, but not in general (the two EPs will be close in parameter space, but won't exactly coincide).

Theory on the lineshape of the spectrum

We also study the underlying factors that influence the lineshape of the output spectrum. To offer a general description, we discuss the coupled microcavities with two waveguide channels, in which we can choose to inject probe light in various forms and measure both the reflection and transmission spectra.

## S6. The reflection and transmission spectra

When we probe the system from port 1, for example, the reflection spectrum takes the form

$$r_1 = 1 - i\frac{\gamma_{c1}\Delta_2}{\Delta_1\Delta_2 - \kappa^2}. \tag{S39}$$

For $\omega_1 = \omega_2 = \omega_0$, it becomes

$$r_1 = \frac{\left(\delta + i\frac{\gamma_1 - \gamma_{c1}}{2}\right)\left(\delta + i\frac{\gamma_2 + \gamma_{c2}}{2}\right) - \kappa^2}{\left(\delta + i\frac{\gamma_1 + \gamma_{c1}}{2}\right)\left(\delta + i\frac{\gamma_2 + \gamma_{c2}}{2}\right) - \kappa^2}. \tag{S40}$$

We thus obtain the absorption spectrum

$$R = |r_1|^2 = \frac{\left(\delta^2 - \frac{(\gamma_1 - \gamma_{c1})(\gamma_2 + \gamma_{c2})}{4} - \kappa^2\right)^2 + \delta^2\left(\frac{\gamma_1 - \gamma_{c1} + \gamma_2 + \gamma_{c2}}{2}\right)^2}{\left(\delta^2 - \frac{(\gamma_1 + \gamma_{c1})(\gamma_2 + \gamma_{c2})}{4} - \kappa^2\right)^2 + \delta^2\left(\frac{\gamma_1 + \gamma_{c1} + \gamma_2 + \gamma_{c2}}{2}\right)^2}. \tag{S41}$$

The denominator vanishes at the poles, while the numerator vanishes at other complex frequencies (zeros if $\gamma_{c2} = 0$). In general cases without PT symmetry, the poles and zeros are not complex conjugate pairs. As a result, when the system is probed by laser light with real frequencies, the reflection spectrum is influenced by both the poles and the zeros, and consequently the dips may not occur at the real part of poles or zeros. This analysis can be readily applied to the situation in which the microcavities are probed by only one waveguide channel, since we can treat the coupling to the second channel as an additional loss to the second cavity and combine $\gamma_{c2}$ and $\gamma_2$ into a total loss rate $\gamma_2'$. In this way, we can write $r_1 \propto \frac{(\delta - \delta_{z1})(\delta - \delta_{z2})}{(\delta - \delta_{p1})(\delta - \delta_{p2})}$, where $\delta_{z1,2} = \omega_{z1,2} - \omega_0$. Thereby, the reflection spectrum can be interpreted as the ratio between two sets of geometric distances in the complex plane: the distances between the point $(\omega, 0)$ and the zeros $(Re(\omega_{z1,2}), Im(\omega_{z1,2}))$, and the distance between the point $(\omega, 0)$ and the poles $(Re(\omega_{p1,2}), Im(\omega_{p1,2}))$. Typically, one can find its local minimum when the probe frequency is close to (not exactly at) $Re(\omega_{z1,2})$ and the local maximum when close to (not exactly at) $Re(\omega_{p1,2})$. Therefore, the split zeros can lead to a doublet in the spectrum. However, it is noted that the poles are further away from the real axis than the zeros ($|Im(\omega_{p1,2})| > |Im(\omega_{z1,2})|$ when $\kappa > \kappa_{th}$ as shown by Fig. 2B in the main text), so that the influence of the zeros on the spectrum is larger than that of the poles. As a result, when $\kappa$ exceeds the $\kappa_{th}$ for the resonant EP, the central transparency window in the spectrum cannot be split by the separated $Re(\omega_{p1,2})$.

On the other hand, when we inject light from port 1 and collect the signal from port 3, the transmission spectrum is

$$|t_{1,2}|^2 = \left| -i\frac{\sqrt{\gamma_{c1}\gamma_{c2}}\kappa}{\Delta_1\Delta_2 - \kappa^2} \right|^2, \tag{S42}$$

which has a constant numerator and reaches maximum when the laser frequency is equal to the real part of the poles. Thus, the transmission spectrum can be utilized to infer the presence of resonant EPs. In experiments, we found the resonant EP by continuously decreasing the intercavity coupling strength $\kappa$ until two peaks in the transmission spectrum $|t_{1,2}|^2$ coalesce.

Now we turn to the scattered output spectrum when the system is probed by the eigenvector of the S-matrix. One eigenvalue of $S$ has the form $\sigma_1 = C_1 \frac{(\omega-\omega_{z1})(\omega-\omega_{z2})}{(\omega-\omega_{p1})(\omega-\omega_{p2})}$ with an eigenvector $v_1$, where $C_1$ is a constant, $\omega_{z1,2}$ are the zeros and $\omega_{p1,2}$ are the poles. When probed by $v_1$, the output signal is $v_{out} = Sv_1 = C_1 \frac{(\omega-\omega_{z1})(\omega-\omega_{z2})}{(\omega-\omega_{p1})(\omega-\omega_{p2})} v_1$, which also vanishes at the complex zeros and diverges at the complex poles. However, under real frequency probe, the output signal does not completely vanish at $Re(\omega_{z1,2})$, and the poles which have different real parts than zeros can shift the location of the local minimum. Thus, the scattered signal spectrum for this eigenvector probe with real frequencies is not able to directly reveal the locations of the zeros and poles. For the other eigenvalue $\lambda_{S2}$ which does not have zeros, the extraction of poles is possible. It is also noted that due to the fact that the poles are associated with each element of the $S$ matrix, there will be no input vector that can avoid the scattering effect of poles. Therefore, the direct detection of zeros will not be straightforward. In our experiments, the locations of zeros are retrieved by curve fitting all the parameters of the system.

## S7. The lineshape of the spectrum at resonant and absorbing EPs

In the Fig. 2 in the main text, we have shown the lineshape of the output spectrum of the coupled microcavities at an absorbing EP displays a doublet instead of a single dip, even though the zeros coalesce. This could be understood as the result of the existence of two poles with different imaginary parts when zeros become degenerate. Here in Fig. S3, we show the output spectrum at a resonant EP, where the intercavity coupling strength is equal to the critical value, i.e., $\kappa = \kappa_{th}$. While two dips can still be observed from the transmission spectrum, the transparency window around the zero detuning is slightly broader than that for the absorbing EP. Note that in a more general situation without the restriction of the EPs, such lineshapes have been previously studied in differentiating electromagnetically induced transparency from Autler–Townes splitting (*28*). Here, the results indicate that the spectrum of the scattered output signal may not be considered reliable for judging EPs, since the peaks or dips may not reflect the exact locations of the resonances (poles) or zeros in the complex plane. It is noted that this is true even in the one-waveguide probing case, where we have chosen the correct eigenvector of the S-matrix as the input optical signal.

It is of interest to explore the sufficient and necessary condition that the absorbing EPs and resonant EPs occur at the same conditions, i.e., the zeros and poles become degenerate simultaneously with the same real part of the frequency. We generalize our discussion to a generic case of two coupled optical modes with two incoming and two outcoming channels, modeled under TCMT. The coupling from mode 1 to mode 2 (vice versa) is described by the coupling rates $\kappa_{12}$ ($\kappa_{21}$). The condition for simultaneous resonant and absorbing EPs can be given by

$$0 = \sqrt{\left(\omega_1 + i\frac{\gamma_{c1} - \gamma_1}{2} - \omega_2 - i\frac{\gamma_{c2} - \gamma_2}{2}\right)^2 + 4\kappa_{12}\kappa_{21}}$$

$$= \sqrt{\left(\omega_1 - \omega_2 - i\frac{\gamma_1 + \gamma_{c1} - \gamma_2 - \gamma_{c2}}{2}\right)^2 + 4\kappa_{12}\kappa_{21}}. \tag{S43}$$

which leads to

$$\left(\omega_1 - \omega_2 + i\frac{\gamma_{c1} + \gamma_2 - \gamma_1 - \gamma_{c2}}{2}\right) = \pm\left(\omega_1 - \omega_2 - i\frac{\gamma_1 + \gamma_{c1} - \gamma_2 - \gamma_{c2}}{2}\right), \tag{S44a}$$

$$0 = \left(\omega_1 + i\frac{\gamma_{c1} - \gamma_1}{2} - \omega_2 - i\frac{\gamma_{c2} - \gamma_2}{2}\right)^2 + 4\kappa_{12}\kappa_{21}. \tag{S44b}$$

It follows that

$$\gamma_{c1} = \gamma_{c2}, \tag{S45a}$$

$$0 = \left(\omega_1 - \omega_2 + i\frac{\gamma_2 - \gamma_1}{2}\right)^2 + 4\kappa_{12}\kappa_{21}. \tag{S45b}$$

Or

$$\gamma_1 = \gamma_2, \tag{S46a}$$

$$\omega_1 = \omega_2, \tag{S46b}$$

$$\left(\frac{\gamma_{c1} - \gamma_{c2}}{2}\right)^2 = 4\kappa_{12}\kappa_{21}, \tag{S46c}$$

Conditions described by Eq. (S45b) or (S46c) can be satisfied by tuning the coupling strengths between the optical modes. Furthermore, we summarize the sufficient and necessary conditions for a resonant EP and an absorbing EP to occur simultaneously: (1) $\gamma_{c1} = \gamma_{c2}$ or (2) $\gamma_1 = \gamma_2$, $\omega_1 = \omega_2$. In another word, **the coalescence of resonant and absorbing EPs happens only at the situation where we have either symmetric coupling channels or two identical optical modes.** This aligns with the theoretical prediction on PT-symmetric systems or gauged PT-symmetric systems (*29*) except that here we treat the system by TCMT. Now we turn to a detailed discussion of each of these cases.

In the first case, we find that the zeros and poles become degenerate simultaneously as long as the coupling channels are symmetric and the intercavity coupling strength is properly tuned, regardless of the other conditions, which enable an absorbing EP and a resonant EP occurring at the same time. This case also represents a gauged PT-symmetric scheme (or PT-symmetric scheme if there is no net gain or loss) where, under TCMT, the scattering EP occurs simultaneously with resonant and absorbing EPs. A special example has been shown by Fig. 3 in the main text, where we have $\omega_1 = \omega_2$ as well, and a coalescence of three types of EPs (resonant, absorbing and scattering EPs) is achieved within the framework of TCMT. In general, the relaxation on the coresonant condition ($\omega_1 = \omega_2$) offers more degrees of freedom for non-Hermitian engineering. For example, with different resonant frequencies for two optical modes and an imaginary coupling strength $\kappa$ (30), anti-PT symmetry (31) can be readily achieved, for which the EPs must be associated with a single-dip lineshape by probing from symmetric coupling channels.

The second case refers to a situation that there are two optical modes with the same resonant frequencies and loss rates. The asymmetric coupling breaks the gauged PT symmetry so that the scattering EP does not occur simultaneously. This case could be found, for example, by coupling two identical microcavities, with a judicious design of the asymmetric coupling channels to achieve EPs. Experimentally it will be not straightforward to engineer such identical modes while maintaining asymmetric coupling.

However, a special case belongs to the simultaneous realization of conditions (1) and (2), bringing complete symmetry to the system. In this case, the EP can only be achieved if either $\kappa_{12}$ or $\kappa_{21}$ vanish. One example can be found in a single microcavity consisting of coupled CW and CCW modes with the suppression of one-side backscattering (32–37), just as shown in the chiral absorber (27). In these cases, the real parts of the resonances and zeros can be derived from the peaks in the absorption spectrum.

Apart from the cases discussed above, the peaks or dips in the reflection spectrum may not reflect the correct positions of the resonances or the zeros. The presence of EPs, especially absorbing EPs, is revealed from the curve fitting of parameters instead of a signature lineshape of the transmission/absorption spectra.

## S8. Lineshape of the spectrum at a CPA EP

CPA EPs are special cases at which the total output signal vanishes at zero detuning due to the degenerate and purely real zeros. To investigate the lineshape of the spectra, we assume $\omega_1 = \omega_2 = \omega_0$, and investigate the scattered output signal $v_{out}$ under the probe of an eigenvector $v_{in}$ of the S-matrix associated with the eigenvalue $\sigma$

$$v_{out} = S v_{in} = \sigma v_{in}, \qquad (S47)$$

where the eigenvalues of S matrix are given by

$$\sigma_{1,2} = \frac{\left(\left(\delta + i\frac{\gamma_1}{2}\right)\left(\delta + i\frac{\gamma_2}{2}\right) + \left(\frac{\gamma_{c1}}{2}\right)\left(\frac{\gamma_{c2}}{2}\right) - \kappa^2\right)}{(\Delta_1 \Delta_2 - \kappa^2)}$$

$$\pm i \frac{1}{(\Delta_1 \Delta_2 - \kappa^2)} \sqrt{\left(\frac{\gamma_{c1}}{2}\left(\delta + i\frac{\gamma_2 + \gamma_{c2}}{2}\right) - \frac{\gamma_{c2}}{2}\left(\delta + i\frac{\gamma_1 + \gamma_{c1}}{2}\right)\right)^2 + \gamma_{c1}\gamma_{c2}\kappa^2}, \qquad (S48)$$

where $\delta = \omega - \omega_0$.

For a one channel CPA EP, $S$ is reduced to a reflection coefficient

$$r_1 = \frac{1}{1 + \frac{i(\gamma_1 + \gamma_2)}{\delta} - \frac{(\gamma_1 + \gamma_2)\gamma_2}{2\delta^2}}. \tag{S49}$$

It follows that

$$R_1 = |r_1|^2 = \frac{1}{1 + \frac{(\gamma_1 + \gamma_2)\gamma_1}{\delta^2} + \frac{(\gamma_1 + \gamma_2)^2 \gamma_2^2}{4\delta^4}} = \frac{\delta^4}{\delta^4 + (\gamma_1 + \gamma_2)\gamma_1 \delta^2 + \frac{(\gamma_1 + \gamma_2)^2 \gamma_2^2}{4}}, \tag{S50}$$

which takes the form of a quartic lineshape, as $R_1 \sim \delta^4$ when $\delta \to 0$.

For a two-channel CPA EP, the quartic lineshape happens only for the generic case, where the waveguide-cavity coupling rates are asymmetric. In this case, the eigenvalues of $S$ are given by

$$\sigma_{1,2} = \frac{\delta^2 + i\frac{\gamma_1 + \gamma_2}{2}\delta - \frac{(\gamma_{c2} - \gamma_{c1})(\gamma_1 - \gamma_{c1})}{4}}{\delta^2 + i(\gamma_1 + \gamma_2)\delta - \frac{\gamma_1 \gamma_{c2} + \gamma_2 \gamma_{c1}}{2}}$$

$$\pm \frac{i\sqrt{\delta^2 \left(\frac{\gamma_{c1} - \gamma_{c2}}{2}\right)^2 - i\frac{(\gamma_{c1} + \gamma_{c2})(\gamma_1 - \gamma_{c1})(\gamma_{c1} - \gamma_{c2})}{4}\delta - \frac{(\gamma_{c1} - \gamma_{c2})^2(\gamma_1 - \gamma_{c1})^2}{16}}}{\delta^2 + i(\gamma_1 + \gamma_2)\delta - \frac{\gamma_1 \gamma_{c2} + \gamma_2 \gamma_{c1}}{2}}. \tag{S51}$$

At $\delta = 0$, one of the eigenvalues approachs zero, which is associated with the perfect absorption channel for the CPA. For $\delta \ll \left|\frac{(\gamma_1 - \gamma_{c1})(\gamma_{c1} - \gamma_{c2})}{4(\gamma_{c1} + \gamma_{c2})}\right|$, we can write this eigenvalue in Taylor expansion. In the case $(\gamma_{c2} - \gamma_{c1})(\gamma_1 - \gamma_{c1}) > 0$,

$$\sigma_1 \approx \frac{\delta^2}{\delta^2 + i(\gamma_1 + \gamma_2)\delta - \frac{\gamma_1 \gamma_{c2} + \gamma_2 \gamma_{c1}}{2}}, \tag{S52}$$

for which we have taken the plus sign in Eq. S49. It is obvious that $|\sigma_1|^2 \sim \delta^4$ when $\delta \to 0$, corresponding to a quartic lineshape in the output spectrum when the eigenvector is chosen as the input. For the case $(\gamma_{c2} - \gamma_{c1})(\gamma_1 - \gamma_{c1}) < 0$, $\sigma_2$ will take the same form as Eq. S50, and similar results will be found.

For the non-generic CPA EP, we cannot do a similar Taylor expansion for the eigenvalue, since $\delta \ll \left|\frac{(\gamma_1 - \gamma_{c1})(\gamma_{c1} - \gamma_{c2})}{4(\gamma_{c1} + \gamma_{c2})}\right|$ is no longer satisfied. Instead, both eigenvalues take the form $\sigma_{1,2} = \frac{\delta^2 + i\frac{\gamma_1 + \gamma_2}{2}\delta}{\left(\delta + i\frac{\gamma_1 + \gamma_2}{2}\right)^2} = \frac{\delta}{\left(\delta + i\frac{\gamma_1 + \gamma_2}{2}\right)}$. Without loss of generality, we consider the example $\gamma_1 > \gamma_{c1}$. The S matrix is given by

$$S = A \begin{pmatrix} S_{11} & -i\gamma_{c1} \\ -i\gamma_{c1} & S_{22} \end{pmatrix}, \tag{S53a}$$

where

$$S_{11} = \left(\frac{\delta}{\frac{\gamma_1 - \gamma_c}{2}} + i\right)\left(\delta + i\frac{\gamma_2 + \gamma_c}{2}\right) - \frac{\gamma_1 - \gamma_c}{2}, \tag{S53b}$$

$$S_{22} = \left(\frac{\delta}{\frac{\gamma_1 - \gamma_c}{2}} - i\right)\left(\delta + i\frac{\gamma_1 + \gamma_c}{2}\right) - \frac{\gamma_1 - \gamma_c}{2}, \quad (S53c)$$

$$A = \frac{\gamma_1 - \gamma_c}{2(\Delta_1\Delta_2 - \kappa^2)} = \frac{\gamma_1 - \gamma_c}{2\left(\delta + i\frac{\gamma_1 + \gamma_2}{2}\right)^2}. \quad (S53d)$$

Furthermore, the eigenchannel now becomes

$$v_{in} = \sqrt{\gamma_c}\begin{pmatrix}-i\\1\end{pmatrix}. \quad (S54)$$

Therefore, the output vector is given by

$$v_{out} = Sv_{in} = \frac{\delta}{\left(\delta + i\frac{\gamma_1 + \gamma_2}{2}\right)}\begin{pmatrix}-1\\-i\end{pmatrix}. \quad (S55)$$

We get the spectrum of the total output power

$$|v_{out}|^2 = \frac{2\delta^2}{\delta^2 + \frac{(\gamma_1 + \gamma_2)^2}{4}}. \quad (S56)$$

We can find that $|v_{out}|^2 \sim \delta^2$ when $\delta \to 0$, which shows the feature of the quadratic lineshape.

In Fig. S4, we show the simulation results of $|r_1|^2$, $|t_1|^2$, $|r_2|^2$ and $|t_2|^2$ at a non-generic CPA EP, which match the experimental results in Figs. 4A and 4B in the main text. Only a single peak is observed in the spectrum of $|t_1|^2$ (or $|t_2|^2$) as a result of the coalescence of the resonant EP and the CPA EP. $|r_2|^2$ still displays two dips due to the influences from both the poles and the zeros. The four elements of the S-matrix become equal at the zero detuning in the case of non-generic CPA EP, as seen in Fig. S4.

In Fig. S5(a), we show the experimental results of the spectra of the output signal under eigenvector probe at the non-generic CPA EP. The output spectra display single absorption dips. As shown in Fig. S5(c), away from the CPA EP, one of the spectra shows a doublet. These are verified by numerical results in Figs. S5(b)(d).

In Fig. S6 and Fig. S7, we show the spectra of the output signal with different relative phases between the two inputs when the system is close to the non-generic CPA EP. It is obvious that the relative phase change can easily split the single dip in one of the output spectra, verifying the importance of the correct relative phase.

**Supplementary figures**

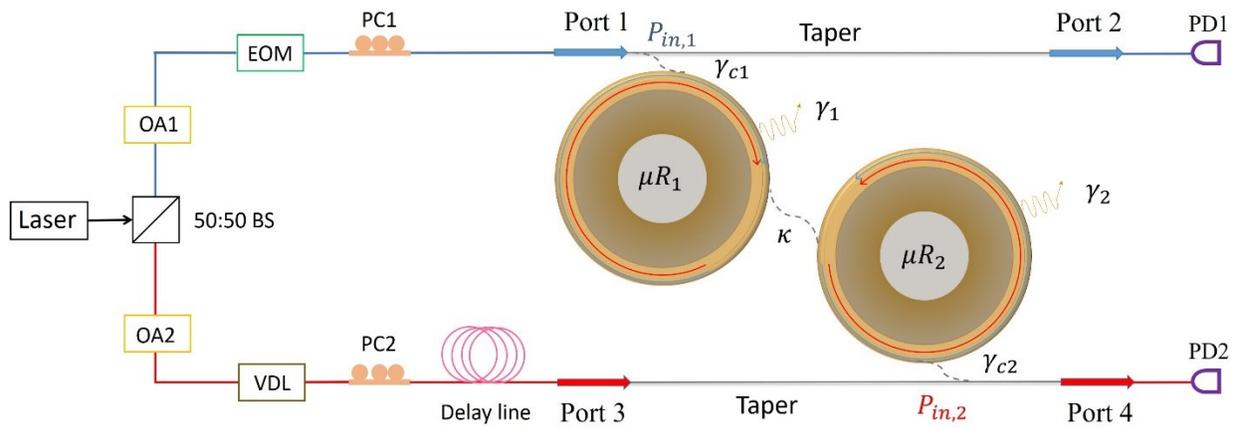

**Fig. S1: Experimental setup for CPA EP measurement with two variable input fields.** BS: beam splitter; OA: optical attenuator; VDL: variable delay line; EOM: electric-optical modulator; PC: polarization controller; PD: photodetector.

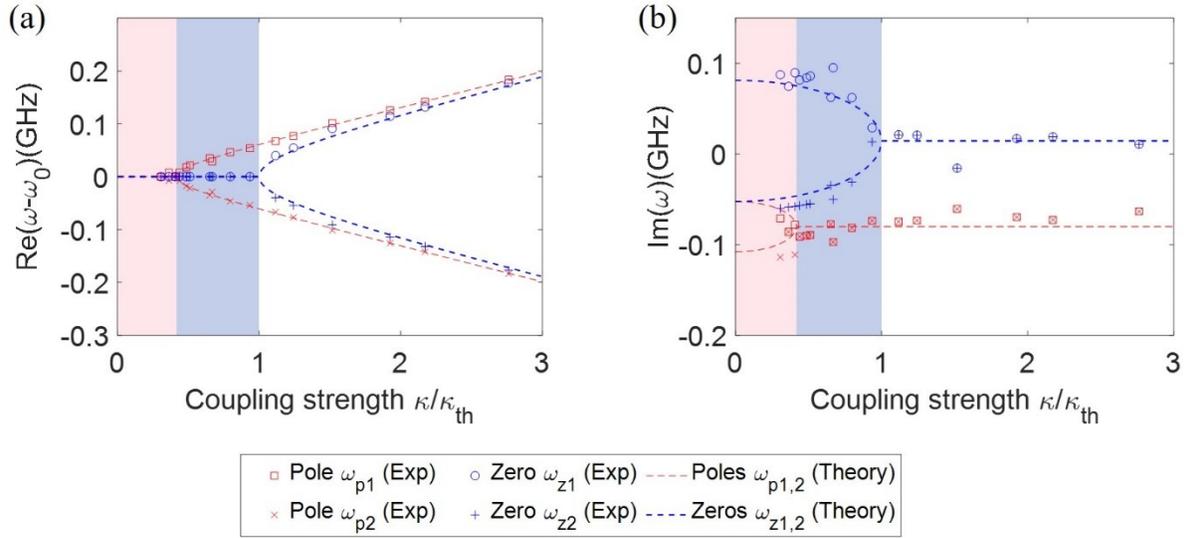

**Fig. S2: Phase transitions of zeros and poles.** Experimentally and theoretically obtained phase transition diagrams for the real parts (a) and imaginary parts (b) of the poles and zeros as a function of normalized coupling strength $\kappa/\kappa_{th}$. In this experiment and simulation, $\gamma_1 < \gamma_2$, which is different from the parameter condition for Fig. 2 in the main text.

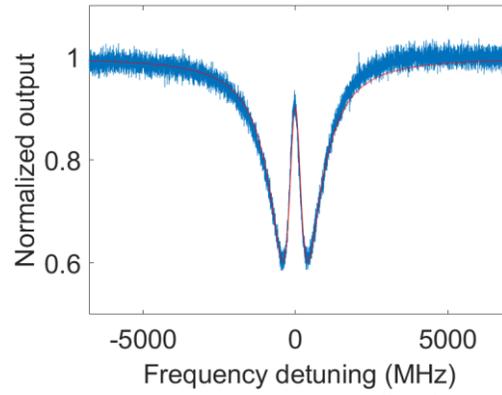

**Fig. S3: Reflection spectrum at a resonant EP.** Reflection spectrum ($|r_1|^2$) of the coupled microcavities at a resonant EP. The lineshape takes a similar form to that of electromagnetically induce transparency, instead of a single broad dip. The blue and red curves are experimental and curve fitting results, respectively.

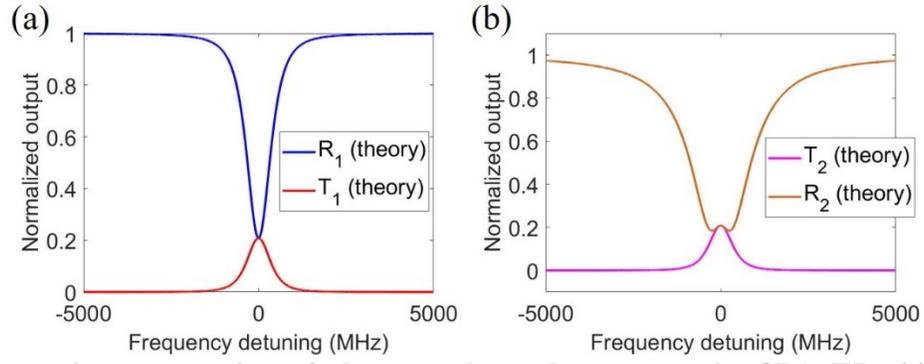

**Fig. S4: Scattering properties of the two-channel non-generic CPA EP.** (a) Spectra of $R_1 = |r_1|^2$ and $T_1 = |t_1|^2$ obtained by TCMT. (b) Simulated spectra of $R_2 = |r_2|^2$ and $T_2 = |t_2|^2$ obtained by TCMT.

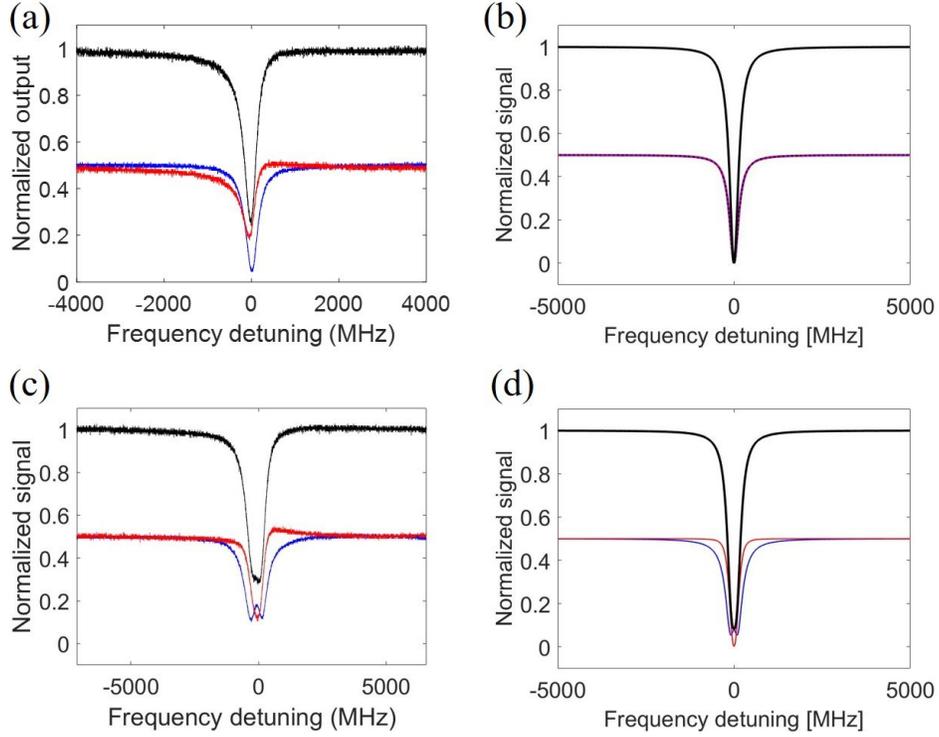

**Fig. S5: Extended data at a CPA EP and a CPA away from the EP.** (a) Experimentally obtained spectra of the output 1 from port 2 (blue curve), output 2 from port 4 (red curve), and the total absorption (black curve) at the non-degenerate CPA EP. The amplitudes of the two input signals at the taper-cavity coupling points are equal, and phases of the two input beams are properly adjusted by the EOM. (b) Simulation result in comparison to the experimental result in (a). Parameters: $\gamma_1 = 64.782 MHz$, $\gamma_2 = 242.93 MHz$, $\gamma_{c1} = 153.86 MHz$, $\gamma_{c2} = 153.86 MHz$, $\kappa = 44.538 MHz$. (c) Experimentally obtained spectra when the system is away from the CPA EP. The amplitudes of the two input beams are equal. (d) Simulation result in comparison to the experimental result in (c). Parameters: $\gamma_1 = 64.782 MHz$, $\gamma_2 = 242.93 MHz$, $\gamma_{c1} = 184.63 MHz$, $\gamma_{c2} = 123.09 MHz$, $\kappa = 95.877 MHz$.

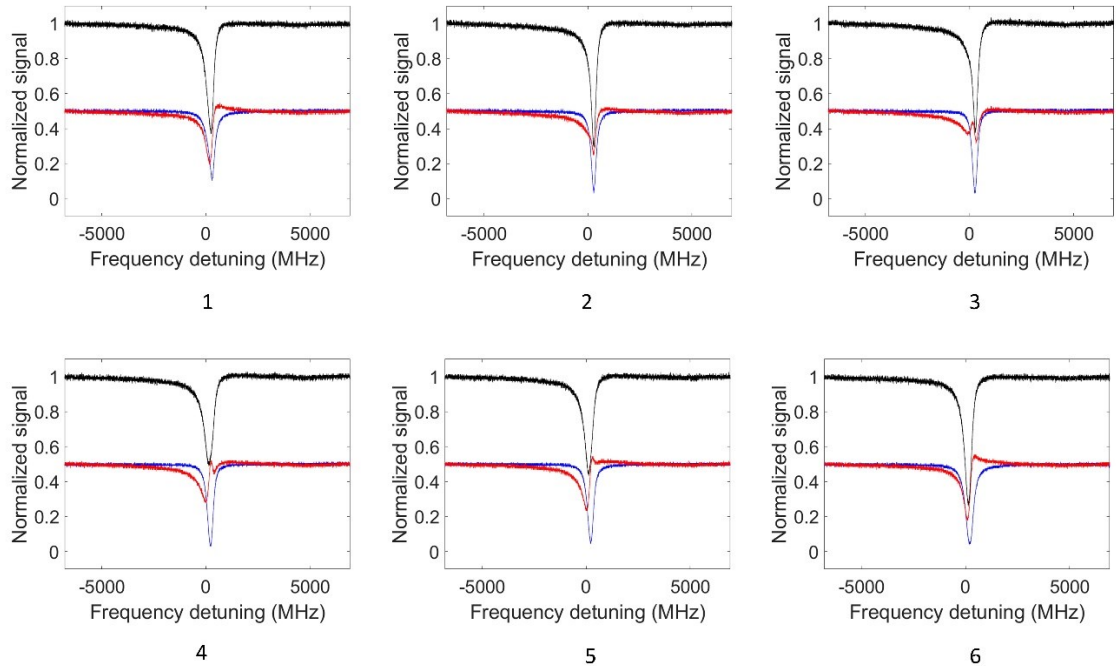

**Fig. S6. Experimental results of the output spectra with different relative phases between the two input fields.** Experimentally obtained spectra of the output 1 from port 2 (blue curve), output 2 from port 4 (red curve), and the total output (black curve) with different relative phases between the two input beams. The amplitudes of the two input beams are equal. The results numbered from 1 to 6 are given in sequence with the change of the relative phase.

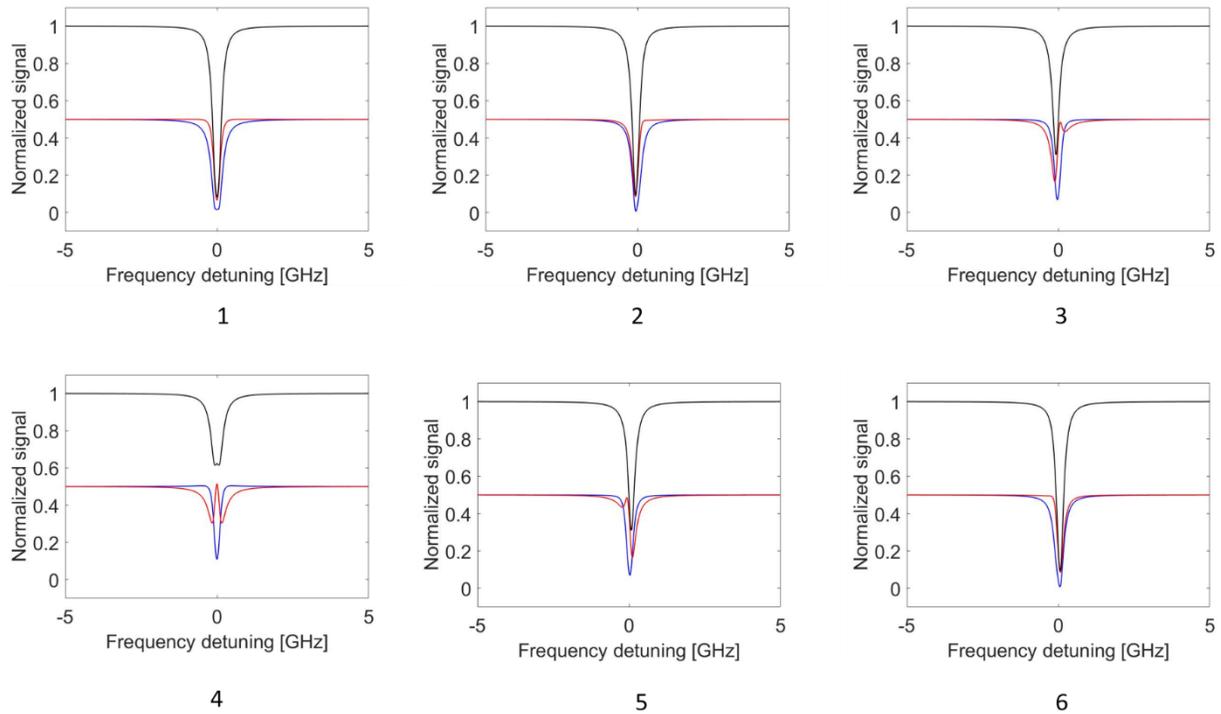

**Fig. S7. Simulation results of the output spectra for different relative phases between the two input fields.** Simulated spectra of the output 1 from port 2 (blue curve), output 2 from port 4 (red curve), and the total output (black curve) with various phase difference between the two input beams. The amplitudes of the two input beams are equal. The results numbered from 1 to 6 correspond to the relative phases equal to $\frac{1}{2}\pi, \frac{1}{6}\pi, -\frac{1}{6}\pi, -\frac{1}{2}\pi, -\frac{5}{6}\pi, -\frac{7}{6}\pi$, respectively.